\documentclass[twocolumn,floatfix]{revtex4}
\usepackage{graphicx}
\usepackage{dcolumn}
\usepackage{amsmath}

\usepackage{amsmath,amsfonts,amssymb}
\usepackage{wrapfig}
\usepackage{graphicx}
\usepackage{bbm}
\usepackage{graphics}

\def \beq {\begin{equation}}
\def \eeq {\end{equation}}
\def \ba {\begin{eqnarray}}
\def \ea {\end{eqnarray}}

\newcommand{\mean}[1]{\langle#1\rangle}

 \newcommand{\Az}{\hat{A}_z}

\begin{document}
\hsize\textwidth\columnwidth\hsize\csname@twocolumnfalse
\endcsname

\title{Measurement of Temporal Correlations of the Overhauser Field\\ in a Double Quantum Dot}
\author{D. J. Reilly$^{1}$, J. M. Taylor$^{2}$, E. A. Laird$^{1}$, J. R. Petta$^{3}$, C. M. Marcus$^1$, M. P. Hanson$^4$ and A. C. Gossard$^4$}
\affiliation{$^1$ Department of Physics, Harvard University, Cambridge, MA 02138, USA}
\affiliation{$^2$ Department of Physics, Massachusetts Institute of Technology, Cambridge, MA 02139, USA}
\affiliation{$^3$ Department of Physics, Princeton University, Princeton, NJ 08544, USA}
\affiliation{$^4$ Department of Materials, University of California, Santa Barbara, California 93106, USA}

\begin{abstract}
In quantum dots made from materials with nonzero nuclear spins, hyperfine coupling creates a fluctuating effective Zeeman field (Overhauser field) felt by electrons, which can be a dominant source of spin qubit decoherence.  We characterize the spectral properties of the fluctuating Overhauser field in a GaAs double quantum dot by measuring correlation functions and power spectra of the rate of singlet-triplet mixing of two separated electrons. Away from zero field, spectral weight is concentrated below 10 Hz, with $\sim 1/f^2$ dependence on frequency, $f$. This is consistent with a model of nuclear spin diffusion, and indicates that decoherence can be largely suppressed by echo techniques.
 \end{abstract}
\maketitle
Electron spins in quantum dots are an attractive candidate for quantum bits (qubits) \cite{Loss_Divencenzo_PRA98,Kane_nature98}. For gate-defined devices made using GaAs, the coupling of single electron spins to  $\sim10^{6}$  thermally excited nuclear spins creates a fluctuating effective Zeeman field (the Overhauser field), ${\bf B}_{nuc}$, with rms amplitude $B_{nuc} \sim$ 1-3 mT  \cite{Erlingsson, Merkulov, Bracker_PRL05, Khaetskii, Paget_PRB77}.  At experimentally accessible temperatures, ${\bf B}_{nuc}$ fluctuates both as a function of position and time, with temporal correlations over a broad range of time scales, and is a dominant source of spin dephasing  \cite{deSousa, Petta_science05, Witzel_PRB06, Koppens_Nature06,Sham_PRB06,Taylor_theory} and low-field spin relaxation \cite{Erlingsson, Merkulov, Bracker_PRL05,Huttel_PRB,Johnson_nature05,Koppens_science05} in these systems. Spin manipulation schemes \cite{Taylor_NatPhys05,Giedke_PRA06,Klauser_PRB06,Witzel_PRB06} to control spin dephasing, such as spin echo and its generalizations, depend critically on a knowledge of correlations and time scales of the fluctuating nuclear environment. 

Previously, fluctuating Overhauser fields have been investigated in atomic systems \cite{Crooker_Nature04} using optical Faraday rotation, superconducting quantum interference devices \cite{clarke} and force-detected magnetic resonance \cite{Rugar}. In quantum dots, dynamic nuclear polarization (DNP) \cite{Abragam,Gammon_Science, Baugh_PRL07,Petta} can drive the nuclear system beyond equilibrium to produce fluctuating currents and feedback effects in connection with Pauli spin-blockade \cite{Ono_PRL04,Koppens_science05, Nazarov_PRL06,Rudner_PRL07}.

In this Letter, we report measurements of the temporal correlations and power spectral densities of the nuclear environment in a two-electron GaAs double-quantum-dot system. In contrast to previous work \cite{Koppens_science05,Baugh_PRL07,Petta}, we do not drive the nuclear system using DNP, but rather probe the statistical fluctuations of the unpolarized nuclear bath in thermal equilibrium \cite{deSousa,Sham_PRB06}. Fluctuations of the Overhauser field are detected as fluctuations in the dephasing time of a two-electron spin state, making use of high-bandwidth proximal charge sensing  \cite{Reilly_APL07}. Fluctuations are found to be broadband over the measurement bandwidth, 40 mHz to 1 kHz, and sensitive to an applied magnetic field in the range $B =$ 0 to 20 mT. Experimental results are shown to be consistent with a simple diffusion model of nuclear dynamics, also presented here. 

\begin{figure}[t!!]
\begin{center}
\includegraphics[width = 8.5cm]{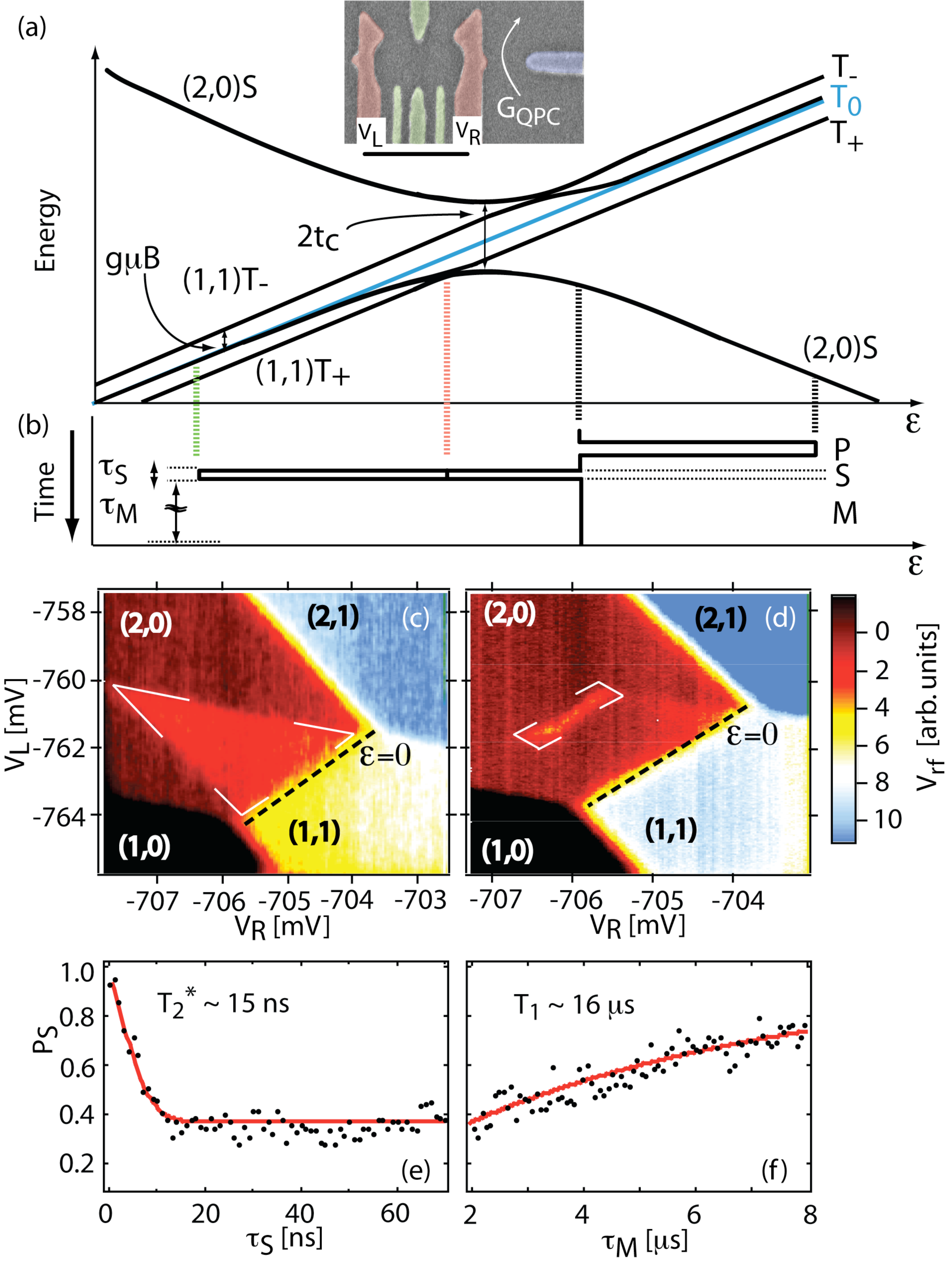}
\caption{(Color online) (a) Schematic energy diagram of the two-electron system. Inset: false-color SEM image of a double-dot with integrated rf-QPC charge sensor similar to the one measured (scale bar is 500 nm). (b) Gate-pulse cycle that is used to prepare (P) the (2,0) singlet, separate (S) into (1,1), either to the $S$-$T_0$ degeneracy (green dashed line) or the $S$-$T_+$ degeneracy (red dashed line), and return to (2,0) for measurement (M).  (c) rf-QPC readout, $V_{\rm rf}$, around the (1,1)-(2,0) transition during application of the cyclic gate-pulse sequence, showing the readout triangle indicated with white lines ($B$ = 0 mT, $\tau_{\rm S}$ = 50 ns). A background plane has been subtracted. (d) $V_{\rm rf}$ as in (c), but for S at the $S$-$T_+$ degeneracy ($B$ = 10 mT). (e) Average value of $P_{S}(\tau_{\rm S})$ at $B$ = 0, $\tau_{\rm M}$ = 2 $\mu$s. Red line is a fit to the theoretical gaussian form. (f) Average value of $P_{S}(\tau_{\rm M})$ showing contrast dependence, $\tau_{\rm S}$ = 50 ns. Red line is a fit to the exponential form (see main text).
}
\vspace{-0.5cm}
\end{center}
\end{figure}

\begin{figure}[t!!]
\begin{center}
\includegraphics[width = 8.5cm]{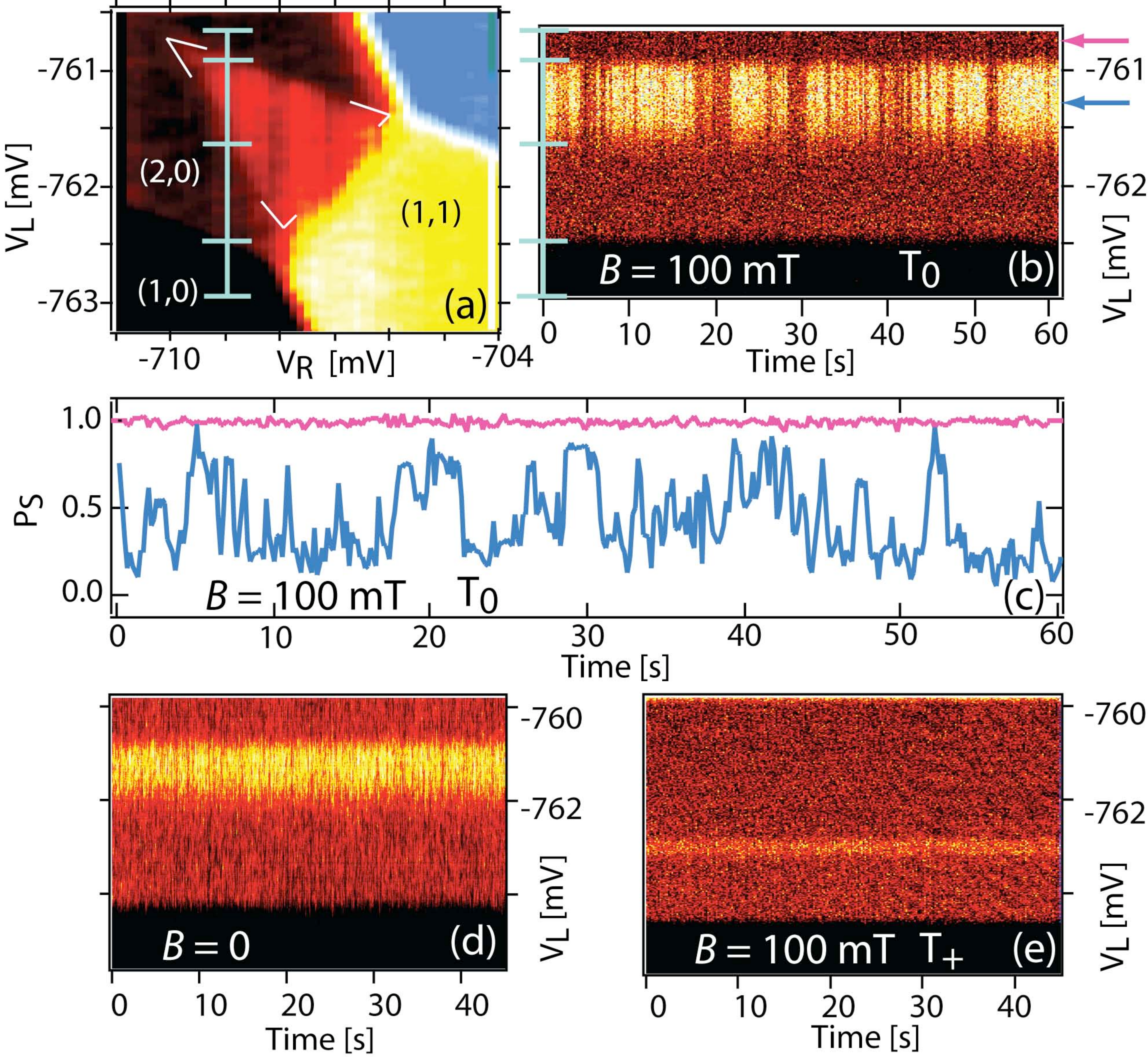}
\caption{(Color online) (a) rf-QPC sensor output $V_{\rm rf}$ as a function of $V_{L}$ and $V_{R}$ with
gate-pulse cycle applied ($\tau_{\rm S}$ = 25 ns,  $\tau_{\rm M}$ = 1.6 $\mu$s,
$B$ = 100 mT). Color scale as in Fig.~1.  (b) Repeated
slices of $V_{L}$ with $V_{R}$ = -709 mV as a function of
time. Markers on left axis correspond to markers
in (a). (c) Sensor output calibrated to $P_{S}$ (blue) along with a measurement of the background QPC noise (pink) from (b) at arrow positions. Bandwidth limited to $\sim$ 3 Hz.
(d) Similar to (b) but for $B$ = 0, color scale same as in Fig.~1. (e) Similar to (b)
but with S-point at $S$-$T_+$ degeneracy, $B$ = 100 mT, color scale same as in Fig.~1.}
\vspace{-0.5cm}
\end{center}
\end{figure}

The double quantum dot is formed by Ti/Au top gates on a GaAs/Al$_{0.3}$Ga$_{0.7}$As heterostructure with a two-dimensional
electron gas (2DEG) with density 2$\times$10$^{15}\, {\rm m}^{-2}$ and mobility
20 m$^{2}$/Vs, 100 nm below the surface (Fig.~1(a), inset), similar to devices reported previously \cite{Petta_science05,Petta}. Measurements are made in a dilution refrigerator with base electron temperature of $\sim$ 120 mK. The conductance $G_{\rm QPC}$ of a proximal radio frequency quantum point contact (rf-QPC) is sensitive to the charge configuration of the double dot. $G_{\rm QPC}$ controls the quality factor of an rf tank circuit, modulating the reflected power of a 220 MHz carrier. Demodulation yields a voltage $V_{\rm rf}$, proportional to $G_{\rm QPC}$,  that constitutes the charge-sensing signal \cite{Reilly_APL07}. The applied field, $B$, is oriented perpendicular to the 2DEG.

Figure~1(a) shows the relevant energy levels of the double dot in the vicinity of the (2,0)-(1,1) charge transition [first (second) index is the charge in the left (right) dot]. Interdot tunneling, $t_{c}$, and detuning, $\epsilon$, from the charge degeneracy are determined by electrostatic gates. A gate-pulse (Fig.~1(b)) cycle prepares new singlets each iteration by configuring the device deep in (2,0), at point (P), where transitions to the ground state singlet, (2,0)$S$, occur rapidly  \cite{Johnson_nature05}. Electrons are then separated to position S in (1,1) for a time  $\tau_{\rm S}$  where precession between the initial singlet and one of the triplet states is driven by  components of the difference in Overhauser fields in the left and right dots, $\Delta {\bf B}_{nuc} = {\bf B}_{nuc}^{l}-{\bf B}_{nuc}^{r}$ \cite{Petta_science05, Taylor_theory}.  

In an applied field, the position of the separation point determines whether the (1,1) singlet ($S$) is nearly degenerate with one of the (1,1) triplets, with which it can then rapidly mix. Mixing of $S$ with $T_{0}$ (the $m_{s}$ = 0 triplet) occurs at large negative $\epsilon$ (green line in Fig.~1(b)) where exchange vanishes. $S-T_{0}$ mixing is driven by components of $\Delta {\bf B}_{nuc}$ {\it along} the total field (applied plus Overhauser fields). In contrast, mixing of $S$ with $T_{+}$ (the $m_{s}$ = +1 triplet), which occurs at a less negative, field-dependent value of $\epsilon$ (red line in Fig.~1(b)) where Zeeman splitting matches exchange, is driven by components of $\Delta {\bf B}_{nuc}$ {\it transverse} to the total field. Measuring the degree of evolution out of the prepared $S$ state following separation, by measuring the return probability to the (2,0) charge configuration after a certain separation time, effectively measures these components of the Overhauser field difference in the two dots. Measurement is carried out by moving the system to position M in (2,0) for a time  $\tau_{\rm M}$ = 5 $\mu$s, during which only $S$ return to (2,0) with appreciable probability. The spin state---triplet or singlet---is thereby converted to a charge state---(1,1) or (2,0), respectively---which is detected by the rf-QPC.  

Figures 1(c,~d) show the time-averaged $V_{\rm rf}$ as a function of gate voltages $V_{L}$ and $V_{R}$. Once calibrated, $V_{\rm rf}$ gives the probability $1 - P_{S}$ that a prepared singlet evolved into a triplet during the separation time $\tau_{\rm S}$.  Inside the readout triangle (see Fig.~1(c)),  triplet states remain blocked in (1,1) for a time $T_{1} \gg \tau_{\rm M}$ \cite{Johnson_nature05}. Similarly, inside the rectangular region indicated in Fig.~1(d), the prepared singlet mixes with $T_{+}$ and becomes blocked in (1,1). Calibration of $V_{\rm rf}$ uses the signal in (2,0) outside the readout triangle, where fast, spin-independent relaxation occurs via (1,0) or (2,1), to define $P_{S} =1$, and the region within (1,1) to define $P_{S} =0$.

Fitting $P_{S}(\tau_{\rm S})$ averaged over tens of seconds with a gaussian \cite{Taylor_theory, Petta_science05} (Fig.~1(e)) gives $T^{*}_{2} = \hbar / (g \mu_{B} B_{nuc}) \sim$15~ns corresponding to $B_{nuc}\sim$ 1.6 mT ($N \sim 6 \times 10^{6}$), where $g\sim -0.4$ is the electron $g$-factor and $\mu_{B}$ is the Bohr magneton. The effect of finite $T_1$ on the calibration of $P_{S}$ can be accounted by introducing a factor $C =  (1-e^{-\tau_{\rm M}/T_{1}}) T_{1}/\tau_{\rm M}$ \cite{Johnson_nature05} that relates $P_{S}$ to the value $P_{S}^{\prime}$ corresponding to infinite $T_{1}$,  $1-P_{S} = (1-P_{S}^{\prime})C$. The dependence of $P_{\rm S}$ on  $\tau_{\rm M}$ (for a fixed $T_{1}\sim$ 16 $\mu$s and  $\tau_{\rm S}$ = 50 ns) is shown in Fig.~1(f). Applying this factor to Fig.~1(e) gives $P_S^{\prime}(\tau_{\rm S} \gg T_2^*) = 1/3$, the expected value \cite{Taylor_theory}, without normalizing the sensor output.
 
With less averaging, $P_{S}$ shows fluctuations that reflect fluctuations of Overhauser field components. Figure 2 shows a slice through the readout triangle, obtained by rastering $V_{L}$ at fixed $V_{R}$, with $B$  = 100~mT, $\tau_{\rm S}$ = 25~ns.  At $B$ = 100 mT, fluctuations in $P_{S}$ have a flickering appearance with broadband time dependence extending to several seconds. Comparing the quieter (pink) trace in Fig.~2c, for point M such that  $(1,1)$ always returns to $(0,2)$, to the fluctuating (blue) trace, where return to $(0,2)$ requires $S-T_0$ mixing by Overhauser fields, we see that the amplitude of the fluctuating signal (blue) is  $\sim$ 100 times larger than the background noise of the charge sensor. At $B$ = 0, slices across the readout triangle does not show a flickering (large, low-frequency) $P_{S}$ signal (Fig.~2(d)). Figure 2(e) shows slices across the $S-T_{+}$ resonance (see Fig.~1(d)).  Here also, $P_{S}$ also does not have a flickering appearance, independent of $B$, reflecting rapid fluctuations of transverse components of $\Delta {\bf B}_{nuc}$. We avoid rapidly cycling through the $S-T_{+}$ transition, which can produce DNP \cite{Petta}.

\begin{figure}[t!!]
\begin{center}
\includegraphics[width = 8.5cm]{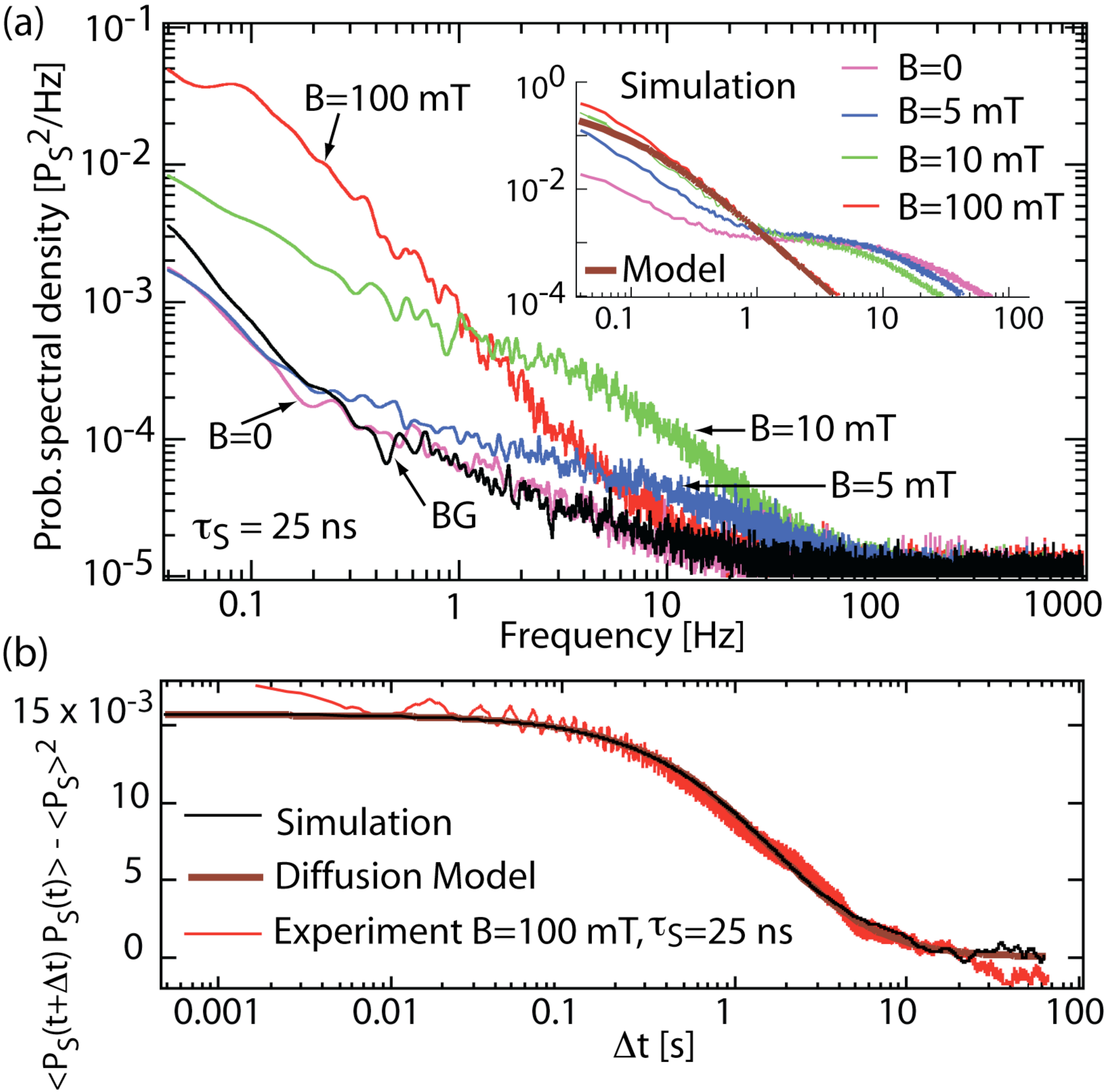}
\caption{(Color online) (a) Power spectra of $P_{S}$ at various magnetic fields, $\tau_{\rm S}$ = 25 ns. Spectra obtained by FFT (with Hamming window) of average of 8 traces sampled at 10 kHz. Background measurement noise (BG) found by setting  $\tau_{\rm S}$ = 1 ns at $B$ = 100 mT. Inset: numerical simulation results for corresponding magnetic fields: $B$ = 0 (pink), $B$ = 5 mT (blue), $B$ = 10 mT (green), $B$ = 100 mT (red).  (b) Autocorrelation $P_{S}$ for $\tau_{\rm S}$ = 25 ns and $B$ = 100 mT (red curve). Model function (Eq.~1) (brown) and Monte Carlo result (black). }
\vspace{-0.5cm}
\end{center}
\end{figure}

To investigate the spectral content of $P_{S}$ fluctuations, fast Fourier transforms (FFTs) of $V_{\rm rf}$ are taken with $V_{L}$ and $V_{R}$ positioned to sample the center of the readout triangle. Figure 3(a) shows power spectra of $P_{S}$,  with $\tau_{\rm S}$  =  25 ns, over the range $B$ = 0 - 100 mT. Measurement at $\tau_{\rm S}$  =  1 ns, where $P_{S} \sim  1$, has a $1/f$ form and is identical to the noise measured outside the readout triangle, and constitutes our background of instrumental noise. At $B=0$ no spectral content above the $1/f$ background noise is seen (Fig.~2(a)). With increasing $B$, an increasing spectral content is observed below $\sim$ 100 Hz. For $B >$ 20 mT, the spectra become independent of  $B$.
The dependence of the power spectrum of $P_S$ on separation time $\tau_{\rm S}$ is shown in Fig.~4. We found that the largest fluctuations over the greatest frequency
range occur for $\tau_{\rm S} \sim T^{*}_{2}\sim$ 15 ns, and these fluctuations show
a roughly $1/f^{2}$ spectrum. Spectra were also obtained out to 100 kHz (not shown) where no additional high frequency components were observed above the background noise. For $\tau_{\rm S} < T^{*}_{2}$,  $P_{S}$ remains near unity with few fluctuations; For $\tau_{\rm S} > T^{*}_{2}$ low-frequency content is suppressed while components in the range $1 - 10$ Hz are enhanced. 

We model fluctuations in $P_{S}$ as arising from the dynamic Overhauser magnetic field in thermal equilibrium. A classical Langevin equation is used to describe fluctuations of $\Delta {\bf B}_{nuc}$ arising from nuclear spin diffusion on distances much larger than the lattice spacing and times much longer than the time-scale set by nuclear dipole-dipole interaction. For $B \gg B_{nuc}$,  correlations of the Overhauser field can be evaluated analytically in terms of a dimensionless operator $\Az^{\beta}$ for each nuclear spin species $\beta$, where $\sum_{\beta} x^{\beta} \hat{A}_{z,l}^{\beta} \equiv {\bf{B}}_{nuc,z}^{l}/B_{nuc}$ and similarly for the right dot, with $x^{\rm ^{75}As } = 1, x^{\rm ^{69}Ga } = 0.6, x^{\rm ^{71}As } = 0.4$.  This gives $\mean{\Az^{\beta}(t+\Delta t) \Az^{\beta}(t)} =  [(1+\Delta t D_{\beta} /\sigma_z^2)^{1/2}(1+\Delta t D_{\beta}/\sigma_{\perp}^2)]^{-1}$, at time difference $\Delta t$, where $D_\beta$ is the species-dependent spin diffusion coefficient, $\sigma_{z}$ is the electron wave function spatial extent perpendicular to the 2DEG (and along the external field) and $\sigma_{\perp}$ is the wave function extent in the plane of the 2DEG, assumed symmetric in the plane. Brackets $\mean{\ldots}$ denote averaging over $t$ and nuclear ensembles. 

Statistics of $P_{S}$ for $S-T_0$ mixing are found using the $z$-component of the Overhauser operators, $\Delta \hat{A}_{z}  =   \sum_{\beta} x^{\beta}(\hat{A}_{z,l}^{\beta} - \hat{A}_{z,r}^{\beta}$). For gaussian fluctuations and a species-independent diffusion constant, $D$, this gives a mean $\mean{P_S}  =  \frac{1}{2} [1 + e^{-2 G^2 \mean{\Delta \hat{A}_{z}^2}}]$ and autocorrelation 
$\mean{P_S(t + \Delta t) P_S(t)} - \mean{P_S}^{2}$
\beq = \frac{e^{-4 G^2 \mean{\Delta \hat{A}_{z}^2}}}{4} \left[ \cosh(4 G^2 \mean{\Delta \hat{A}_{z}(t+\Delta t) \Delta \hat{A}_{z}(t)}) - 1 \right],
 \eeq
where $G =  \tau_{S}/T_{2}^{*}$ is a gain coefficient. The autocorrelation function at $B$ = 100 mT shown in Fig. 3(b) is obtained by Fourier transforming the power spectrum \cite{footnote_neg_cor}. We fit to the autocorrelation function using a contrast factor, $C$, (see Fig. 1(f) and discussion), and the diffusion coefficient, $D$, as fitting parameters. Wavefunction widths are taken from numerical simulations of the device \cite{Stopa}, $\sigma_{z}$  =  7.5 nm,  $\sigma_{\perp}$  =  40 nm. The fit gives $D \sim 10^{-13}$~cm$^2$/s, consistent with previous measurements on bulk GaAs samples using optical techniques  \cite{Paget_PRB77}. In Eq. (1) the dependence on $\tau_S$ leads to a scaling of the correlation time of $P_S$ by $G^2$ to find the underlying Overhauser field correlation time. For fields $B >$ 20 mT, the data in Fig.~3(b) indicate an autocorrelation time of $\sim$ 3 s for $P_{S}$ corresponding to a time $\tau_{d} \sim$ 10 s for $\Delta A_{z}$ to decorrelate by half of its initial value.
\begin{figure}
\begin{center}
\includegraphics[width = 8.5cm]{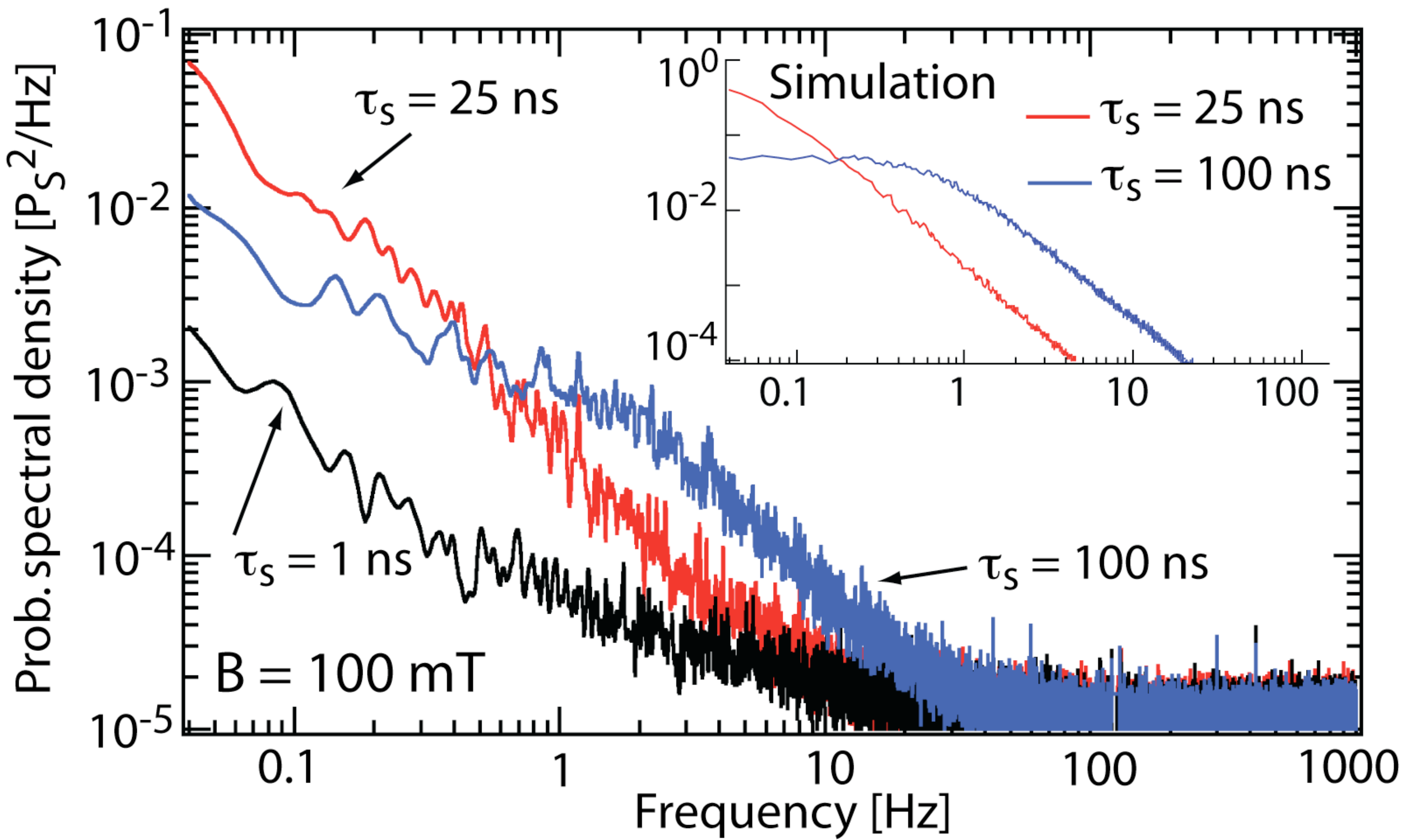}
\caption{(Color online)  Power spectra of $P_S$ at $B$ = 100 mT for separation times $\tau_{\rm S}$ = 25 ns (red) and $\tau_{\rm S}$ = 100 ns (blue).  Setting $\tau_{\rm S}$ = 1 ns (black) yields background noise. Inset shows simulation results for $B$ = 100 mT, $\tau_{\rm S}$ = 25 ns (red) and $\tau_{\rm S}$ = 100 ns (blue).  Note the suppression of low-frequency content and enhancement of mid-frequency content for long $\tau_{\rm S}$ in the experiment and simulation. }
\vspace{-1.0cm}
\end{center}
\end{figure}

Near $B \sim 0$, transverse components of the nuclear field lead to rapid dephasing of nuclear spins. In this regime, we use a Monte Carlo method to simulate nuclear dynamics \cite{Taylor_unpublished}. Figure 3(b) shows that numerical and analytical approaches agree at higher fields, where both methods are applicable. Numerical power spectra for $B \sim 0$ are shown in the inset of Fig. 4. 

Experiment and theory show reduced low-frequency spectral content as $B$ decreases toward zero. This can be understood as arising from the influence of the transverse nuclear fields at low $B$, which rapidly dephase nuclear spins and suppress long time correlations in $\Delta B_{nuc}$. Similar behavior, though independent of $B$,  is observed in the spectra of $P_{S}$ at the $S - T_{+}$ resonance (not shown). Below $B \sim$ 10 mT, an increased spectral content at  frequencies between 1 - 10 Hz is observed in the experiment and theory. The frequency at which the spectra intersect however, remains constant ($\sim$ 1 Hz) in the simulations but increases at low $B$ in the experimental data. We are able to approximate this behavior in the simulation by increasing the diffusion coefficient ($D \sim 10^{-12}$ cm$^2$/s at $B$ = 0), implying an enhancement of diffusion, beyond typical values \cite{Paget_PRB77}, as $B$ approaches zero. This may be due to the growing influence of non-secular terms in the dipole-dipole interaction at low magnetic field  \cite{Abragam,deSousa}. Diffusion maybe further enhanced at low $B$ as a result of electron mediated flip-flop of nuclear spins \cite{Sham_PRB06,Xuedong_PRB06}, an effect neglected in the simulation.

Finally, we model how the separation time for the two-electron spin state affects the power spectra.  Simulated spectra are shown in the inset of Fig.~4 for $\tau_{\rm S}$ = 1 ns, 25 ns and 100 ns  at $B$ = 100 mT. Good agreement with experiment is achieved when again accounting for the additional $1/f$ noise and contrast reduction. We find that $\tau_{\rm S}$ acts to filter fluctuations in $\Delta B_{nuc}$, so that for $\tau_{S} \gg T_{2}^{*}$, low frequency  correlations in $\Delta B_{nuc}$ are suppressed in the spectra of $P_{S}$ (see Eq.~1). This filtering effect leads to the turn-over at $\sim$ 2 Hz evident in the spectra for $\tau_{\rm S}$ = 100 ns. For $\tau_{S}\sim T_{2}^{*}$, little filtering occurs and the power spectra of $P_{S}$ reflect the underlying intrinsic fluctuations of the Overhauser magnetic field. 

We thank  L. DiCarlo, A. C. Johnson, and M. Stopa for contributions. This work was supported by DARPA, ARO/IARPA, NSF-NIRT (EIA-0210736) and Harvard Center for Nanoscale Systems. Research at UCSB supported in part by QuEST, an NSF Center.
\small

\end{document}